\documentclass[11pt,a4paper]{article}

\usepackage{endnotes}
\usepackage{ulem}
\usepackage{color}
\let\footnote=\endnote

\newcommand{\vct}[1]{\boldsymbol{#1}}
\newcommand{\mtrix}[3]{\langle \,#1\,|\,#2\,|\,#3\,\rangle}
\newcommand{\avr}[1]{\langle \,#1\,\rangle}
\usepackage{amsmath, amssymb}
\usepackage{bm}

\usepackage[dvipdfmx]{graphicx}
\usepackage{amsmath, amssymb}

\begin{document}

\begin{center}

{\Large
 The neutron skin-thickness of $^{208}$Pb determined by
electron and proton scattering 
}

 \vspace{5mm}
 \noindent
 Toshio Suzuki, Rika Danjo and Toshimi Suda
 
Research Center for Accelerator and Radioisotope Science,
Tohoku University \\Sendai 982-0826, Japan

\end{center}

\vspace{3mm}
\noindent
email: kt.suzuki2th@gmail.com

\vspace{5mm}
\noindent
Abstract:
Electron as well as proton elastic scattering is not able to determine
the point proton and point neutron densities, $\rho_\tau(r), (\tau=p,n)$,
separately.
If both scatterings are analyzed consistently, those densities
would be determined uniquely, since 
the two densities are observed by different combinations from each other.
The previous experiments did not provide $\rho_\tau(r)$ uniquely,
but the values of the mean square radii of  $\rho_p(r)$, $\avr{r^2}_p$,
and  of $\rho_n(r)$, $\avr{r^2}_n$, 
are shown to be determined consistently through the fourth moment of the observed
charge density, $\avr{r^4}_c$,  in $^{208}$Pb.
The previous analyses of $(\gamma, \pi^0)$ and $\bar{p}$-nucleus
obtained a similar value of $\avr{r^2}_n$, but they do not yield
the experimental value of $\avr{r^4}_c$ observed in electron scattering.




\vspace{1cm}

One of the long-standing problems in the nuclear physics 
is if the mean square radius of the point-neutron distribution(n-msr)
\footnote{The abbreviation
of the `rms'(root mean square)-radius is frequently used in the literature,
but it is convenient for the present purpose to employ `msr' for
the mean square radius, because electron scattering observes the value of the msr
($\avr{r^2}_c$), together with the higher moments, $\avr{r^{2n}}_c, (n=2, 3, \cdots)$,
rather than the square root of the msr.}
 in nuclei is determined experimentally.
 On the one hand, the msr of the point proton distribution(p-msr)  has been determined
 almost model-independently through the probe with well-known electromagnetic interaction like
 electron scattering\cite{bd,deforest}, and has played an essential role for understanding
 nuclear structure from the beginning of nuclear physics history\cite{bm}.
 On the other hand, its counter-part, the n-msr, has been still under discussions
 experimentally and theoretically, because its experimental study has to employ probes
 with strong interaction.
 Compared with electromagnetic interaction,
 the understanding of the strong interaction in the many-nucleon system
is not known without ambiguity and, moreover, the reaction  mechanism
with strong interaction is not yet well established,
as in proton scattering\cite{ray78,ray79,star,mark,sak,brown,zen}.
 As a result, there is no data table
of the n-msr yet, unlike those of the p-msr throughout the periodic table\cite{vries,ang}.

Both electron and proton as probes
interact with the protons and neutrons
in the nucleus,
but each scattering itself is not able to observe the point-proton distribution, $\rho_p(r)$
and the point-neutron distribution, $\rho_n(r)$, separately.
Electron scattering observes the nuclear charge density, $\rho_c(r)$, as\cite{deforest}
\begin{equation}
\rho_c(r)=\rho_{cp}(r)+\rho_{cn}(r)\label{cden},
\end{equation}
where $\rho_{c\tau}(r), (\tau=p,\,n)$ denote the $\rho_{\tau}(r)$ 
folded by a single
proton and neutron charge distributions, respectively,
which are supposed to be known in other experiments. 
Proton scattering is analyzed by assuming, for example, the potential as\cite{mark}  
 \begin{equation}
U(r)=\sum_\tau \int d^3r'\rho_\tau(r)t_\tau(|\vct{r}-\vct{r}'), \label{ps}
\end{equation}
where $t_\tau$ indicate the free nucleon-nucleon $t-$matrix.
Thus, the electron and proton are scattered by $\rho_\tau(r)$ 
in different ways.
In principle, therefore, if proton scattering were described transparently as electron
scattering, the combined data would provide almost uniquely $\rho_\tau(r)$.

Ref.\cite{ray78, ray79,hoff,star} aimed to analyze electron and proton scattering consistently.
In the first step, the author of Ref.\cite{ray79} obtained $\rho_\tau$  
by using experimental values of the right-hand side in Eq.(\ref{cden}),
but assuming each contribution
of $\rho_\tau(r)$ to it model-dependently\cite{ber},
because electron scattering can not observe
them separately as mentioned before.
Next, proton scattering is analyzed with the use of the obtained
$\rho_{\tau}(r)$, and he determines the best $\rho_{\tau}(r)$
to reproduce the proton-scattering cross sections.
Third, the obtained new $\rho_{\tau}(r)$ is examined
if the original electron scattering data are reproduced.
According to such iterations, it is found that a few repetitions are enough
for the convergence, if the first trial function of $\rho_{\tau}(r)$ 
are well prepared\cite{ray79}. The model-dependence in the first step is expected to
disappear in the iterations.

Nevertheless, various experiments and their analyses\cite{jlab, cry1,cry2,di2}
together with proton scattering\cite{ray78,ray79,hoff,star,mark,sak,brown,zen}
were repeated for the  last forty years.
The reason may be because
there is no unique understanding of the reaction mechanism 
and used parameters, so that
there is not enough confidence for $\rho_n(r)$ to be derived 
in the analyses of the proton scattering.
Indeed, all the analyses in Refs.\cite{ray78,ray79,hoff,star,mark,brown,sak,zen}  
used different descriptions for the strong interaction with $\rho_\tau(r)$,
sometimes with physically different parameters like assumed effective masses of nucleons
and mesons in the nuclear medium as in Refs.\cite{star,mark, brown} and \cite{sak,zen}.

Among experiments accumulated until now, however, the noteworthy fact is that,
when the $\rho_p(r)$ are inferred from the electron-scattering
data, all the analyses of the proton scattering\cite{ray78,ray79,hoff,star,mark,brown,sak,zen}
predict the value of
the neutron skin thickness, $\delta R$, to be about $0.14\sim 0.2$,
where $\delta R$ is defined by $\delta R=\sqrt{\avr{r^2}_n}-\sqrt{\avr{r^2}_p}$ 
with the n-msr and p-msrh denoted as $\avr{r^2}_n$ and $\avr{r^2}_p$,
respectively.
This fact implies that the value of 
$\avr{r^2}_n$ proportional to $\int r^4\rho_n(r)$dr depends weakly
on the various reaction mechanisms assumed
and different parameters used in the proton-scattering analyses,
as far as the unfolded charge densities inferred from electron scattering are employed.
In order to confirm the reliability of the value of $\avr{r^2}_n$, therefore, 
it may not be appropriate to compare the $\rho_n(r)$-profiles with one another,
within proton-scattering analyses.
Instead, it is better to explore a bridge 
between electron- and proton-scattering analyses, to confirm the validity of $\avr{r^2}_n$.
 It will be shown below that the moments of the charge distribution observed in electron
scattering play a role of such a bridge.

Fortunately, the second moment(msr) of $\rho_c(r)$, $\avr{r^2}_c$,
does not depend on $\rho_{cn}(r)$\cite{ks1}.
The p-msr is given by the second moment
as\cite{kss}
\begin{equation}
\avr{r^2}_c
 =\frac{1}{Z}\mtrix{0}{\sum_{k=1}^Z r_k^2}{0}+r^2_p+r^2_n\frac{N}{Z}+C_{\rm rel}, \label{2ndc}
\end{equation}
where the first term of the right-hand side stands for the p-msr, $\avr{r^2}_p$, while 
$r^2_p$ and $r^2_n$ indicate the mean square radii of the proton and neutron themselves,
respectively\cite{kss}.
Ref.\cite{kss} used their values as $r_p=0.877$ fm and $r^2_n=-0.116$ fm$^2$. 
The last term of Eq.(\ref{2ndc}) represents the relativistic correction 
up to order of $1/M^2$ which is written as\cite{kss}
\begin{equation}
C_{\rm rel}= \frac{1}{M^2}\left(\frac{1}{Z}\sum_{k=1}^A\mu_k
 \mtrix{0}{\vct{\ell}_k\!\cdot\!\vct{\sigma}_k}{0}+\frac{3}{4}
 +\frac{1}{2Z}\sum_{k=1}^Z\mtrix{0}
 {\vct{\ell}_k\!\cdot\!\vct{\sigma}_k}{0} \right),\label{r}
\end{equation}
with $M$ denoting the nucleon mass to be $939$ MeV, and $\mu_k$ the anomalous
 magnetic moment, $\mu_p=1.793$ for proton and $\mu_n=-1.913$ for neutron.
Estimation of the relativistic correction is almost model-independent in $^{208}$Pb
as a double closed shell nucleus.
For example, 
Ref.\cite{kss} estimated, against the experimental value
 of $\avr{r^2}_c=30.283(0.154)$ fm$^2$\cite{vries,ang},
 the value of $C_{\rm rel}$ to be about $0.021$ fm$^2$ 
with the use of the Skyrme mean-field models, yielding $\avr{r^2}_p=29.671$ fm$^2$.
We note that the last $\mu_p-$independent term in the parenthesis
of the right-hand side in Eq.(\ref{r})
stems from the  the Foldy-Wouthuysen transformation\cite{bd}
of the four-component wave function to the
two-component one, together with the first two terms\cite{kss}.

\begin{table}
 \hspace*{0cm}%
\begin{tabular}{|l|c|c|c|c|c|c|c|} \hline
$\,$&
$\avr{r^4}_c$&
$\avr{r^2}_c$&
$\avr{r^4}_p$&
$\avr{r^2}_p$&
$\avr{r^2}_n$&
$\Delta_2/\Delta_4$&
 $\delta R$
\\ \hline
\rule{0pt}{12pt}%
$(e, N)$  & $1171.981$ &$30.283$& $ \,$ &$ (29.671)$ & $\,$ &$(\Delta_2=0.612)$ & $\,$\\
$(p, N)$  & $ [1173.3]$ &$[30.265]$& $ 1119.6$ &$29.790$ & $31.900$ &$(0.475/3.416)$&$0.198$\\ 
\,LSA & $[1171.981]$ & $[30.283]$ & $1111.855$ &$29.738$&$31.507$ &$\Delta_4=2.605$  & $0.160$\\
$(\bar{p}, N)$ & $ (1156.047)$ &$\,$& $(1098.016)$ &$(29.554)$ & $31.311$ &$\,$&$0.159$ \\
$(\gamma,\pi^0)$& $(1155.040)$ & $\,$&$(1096.854)$ & $(29.569)$& $31.114$ & $\,$&$0.140$\\
 \hline
\end{tabular}
\caption{
The $n$th moments of the charge, proton, and neutron distribution in $^{208}$Pb obtained
 in various analyses.
The values in the square brackets indicate the experimental values used in the analyses,
 while those in the parentheses are calculated with the assumed distributions.
   The values of LSA are obtained by the least
 squares analysis of the non-relativistic mean-field models
 in Ref.\cite{kss}.
 The neutron skin-thickness, $\delta R$, is defined by $(\sqrt{\avr{r^2}_n}-\sqrt{\avr{r^2}_p})$.
  All the values are given in units of fm$^n$. 
   For the details, see the text.  
}
 \label{mom}
\end{table}

The $n(\ge 4)$th moment of the charge distribution depends not only
on the $m(\le n)$th moments of $\rho_p(r)$\cite{ks1},
but also on the $m(\le (n-2))$th ones of $\rho_n(r)$. 
For example, $\avr{r^4}_c$ is given by\cite{kss}
\begin{equation}
\avr{r^4}_c =\avr{r^4}_p+\frac{10}{3}r_p^2\avr{r^2}_p
+\frac{10}{3}r_n^2\avr{r^2}_n\frac{N}{Z}+\Delta_4, \label{4thmc}
\end{equation}
where $\Delta_4$ represents the fourth moment of a single proton and neutron
charge distribution and relativistic corrections.
Ref.\cite{ks1,kss} shows the explicit expression of $\Delta_4$ and its value 
is estimated model-dependently in Ref.\cite{kss}.
The last three terms of Eq.(\ref{2ndc}) will be expressed as $\Delta_2$ from now on
in the same way.

On the one hand, in electron scattering, the experimental value of
$\avr{r^4}_c$ is known\cite{kss,emrich}
and the second term of the right-hand side of the above equation
is determined by Eq.(\ref{2ndc}), taking into account a small correction due to $\Delta_2$.
On the other hand, in proton scattering,
Ref.\cite{mark} summarized the values of the moments 
in the first three values of the right-hand side,
which were obtained from $\rho_\tau(r)$ derived
in their analyses of the experimental data in Ref.\cite{ray78,ray79, hoff,star}.
Ref.\cite{mark} listed also the value of the moments of the charge distribution
obtained in the analysis of electron-scattering data\cite{vries2}.
These values are listed in the $(e,N)-$ and $(p,N)-$rows of Table \ref{mom}
in electron and proton scatterings, respectively.

The experimental values of 
$\avr{r^4}_c=1171.981$ fm$^4$
and $\avr{r^2}_c=30.283$ fm$^2$ in the $(e,N)$-row are obtained in the
sum-of-Gaussians analysis of electron-scattering data\cite{vries,emrich},
which are almost equal
to those by the Fourier-Bessel one\cite{kss,ts1},
while Ref.\cite{mark} employs the values taken from
Ref.\cite{vries2} which assumed the three-point Gaussian distribution
for the charge density.
The values of $r_p$ and $r_n^2$ in Ref.\cite{ray79} cited in Ref.\cite{mark},
are taken from Ref.\cite{holer} which provides $0.836$ fm and $-0.117$ fm$^2$, respectively.
By employing these values, in order to reproduce the experimental value
of $\avr{r^4}_c=1173.3$ fm$^4$, we need the value of $\Delta_2=0.475$ fm$^2$
and $\Delta_4=3.416$ fm$^4$, which are comparable to  
$\Delta_2=0.612$ fm$^2$ and $\Delta_4=2.605$ fm$^4$ listed in the $(e,N)$- and LSA-rows
in Table \ref{mom}, respectively. 
These values of LSA have been obtained by the least squares analysis(LSA)
of the non-relativistic mean-field(NMF)-models in Ref.\cite{kss}.

The least squares method is as follows.
First, a number of the NMF-models are chosen arbitrarily in the literature\cite{stone},
 where there are more than $100$ versions accumulated for the last $50$ years.
By using them,
the values of $\avr{r^4}_c$ and $\avr{r^2}_p$ are calculated according
to Eqs.(\ref{2ndc}) and (\ref{4thmc}).
Next, those values are plotted in the $\avr{r^2}_p-\avr{r^4}_c$ plane,
to obtain the least squares regression line(LSL)
between  $\avr{r^4}_c$ and $\avr{r^2}_p$.    
Finally, The value of $\avr{r^2}_p$ accepted for the used model-framework
is determined by  the cross point(LSL-value) of the LSL
and the line of the experimental value of $\avr{r^4}_c$.
Ref.\cite{ts1} proved that the LSL-values provides uniquely the value of each component of
the reference formula like Eqs.(\ref{2ndc}) and (\ref{4thmc}), as listed in Table \ref{mom}.
It should be noted that these values of the components are not experimental values,
but are allowed in the model-framework as for the NMF-models\cite{ts1}.
The relativistic mean-field(RMF)-models provide different LSL-values\cite{kss} from those
 in Table \ref{mom}, because the definitions of $\avr{r^n}_c$ and $\avr{r^n}_\tau$
 themselves are different from Eqs.(\ref{2ndc}) and (\ref{4thmc}).
The LSL-values obtained for the RMF-models
 which should be compared with the values obtained
 with the relativistic analysis of proton scattering,
 including the relativistic spin-orbit charge density\cite{ks2} which is more important
 for $\Delta_4$ than that in Table \ref{mom}\cite{kss}.
As far as the authors know, however, there is not such a analysis which is performed consistently
 with electron-scattering data by iteration-method\cite{sak2}.
 .

Now, since proton\cite{ray78,ray79,hoff,star,mark}- and electron\cite{vries,vries2}-scatterings
were analyzed by using the same $\rho_\tau$, the experimental values of $\avr{r^4}_c$ and $\avr{r^2}_c$
observed in electron scattering should be reproduced by the values of their components
determined by proton-scattering analyses,  
according to Eqs.(\ref{2ndc}) and (\ref{4thmc}). Table \ref{mom} shows that the values of
the components in the $(p,N)$-row reproduce remarkably well the experimental values
of $\avr{r^4}_c$ and $\avr{r^2}_c$. 
We note that in order to obtain $\avr{r^4}_c=1173.3$, Ref.\cite{mark} published in 1995
did not use the right-hand side of Eq.(\ref{4thmc}) in Ref.\cite{ks1} which appeared in 2019.

There are a few comments on Table \ref{mom}.
First, the values of the components listed in the $(p,N)$-row are in good agreement with the LSL-values.
The only exception is between the values of $\avr{r^4}_p$.
The value given by proton scattering is $1119. 6$ fm$^4$, while that by the LSA $1111.855$ fm$^4$.
In electron scattering, however, the observed $\avr{r^4}_c$ depends on the values of $\avr{r^4}_p$,
$\avr{r^2}_p$ and $r^2_p$ as
$\avr{r^4}_p+\frac{10}{3}r_p^2\avr{r^2}_p$ in Eq.(\ref{4thmc}).
In taking into account the value of $r_p=0.836$ fm used in the proton-scattering analyses
and $0.877$ fm in the LSA, the above sums become $1188.096$ fm$^4$ and $1189.000$ fm$^4$
in the proton scattering analysis and the LSA, respectively.
Thus, it is seen that electron scattering provides a strong constraint on the values of
 the moments of $\rho_p(r)$.

Second, Eq.(\ref{4thmc}) is written as
\begin{equation}
\avr{r^4}_c =\avr{r^4}_p+\frac{10}{3}(r_p^2+r^2_n\frac{n}{Z})\avr{r^2}_p
+\frac{10}{3}r_n^2\frac{N}{Z}(\avr{r^2}_n-\avr{r^2}_p)+\Delta_4 \label{4thmcc}
\end{equation}
The Table $\ref{mom}$ provides for (p-N) and LSA, respectively, as
\begin{align}
1173.3&=1119.6+51.548-1.264+3.416,\\
 1171.981&=1111.855+58.572-1.051+2.605,
\end{align}
where each value in the right-hand side corresponds to those in Eq.(\ref{4thmcc}). 
As mentioned in the previous comment, the sum of the first two terms of the above equations 
are constrained by $\rho_p(r)$. The last term denotes the value of $\Delta_4$ which
depends on the fourth moment of the nucleon and the spin-orbit density\cite{ks1,kss}.
Thus, the contribution of $(\avr{r^2}_n-\avr{r^2}_p)$-term to $\avr{r^4}_c$ is about $0.1\%$,
reflecting $|r^2_n|<r^2_p$.
Hence, in electron scattering only, it seems to be impossible to determine the value of $\delta R$.
In contrast to Eq.(\ref{4thmcc}) in electron scattering, Eq.(\ref{ps}) indicates
\begin{equation}
t_p\rho_p+t_n\rho_n=(t_p+t_n)\rho_p+t_n(\rho_n-\rho_p),
\end{equation}
implying that the last term contributes much to proton-scattering cross section.
Indeed, for example, Ref. \cite{hoff} pointed out that $\rho_n(r)$,
which provides $\delta R>0$,
is sensitive to the maximum- and minimum-position and the slope of the angular distribution
in proton-scattering cross section.

Third, Table \ref{mom} lists the values of $\avr{r^2}_n$
 observed in $\bar{p}-$atom as $(\bar{p},N)$\cite{cry1},
 and in coherent $\pi$ production as $(\gamma, \pi^0)$\cite{cry2}
 for reference. Their values were obtained by using the two-point
 Fermi distributions of $\rho_\tau(r)$.
 When using the given two parameters, and $r_p=0.877$ fm and $r^2_n=-0.116$
fm$^2$,
the values of $\avr{r^4}_c$ are obtained as in Table \ref{mom}
\footnote{Refs.\cite{cry1} and \cite{cry2} provided the values of $\delta R$
to be $0.16$ and$0.15$ fm, respectively, obtained approximately by the two-point Fermi distributions.
Table \ref{mom} lists the values by the exact formulae with the same two parameters for the
Fermi distribution, because the values of the fourth moments are required for comparison.}.
They are much smaller than others, implying that consistent analyses
are necessary for discussions of $\delta R$.

Finally,
all the arguments in the present paper have not taken into account the
experimental errors, because our purpose is to understand a qualitative
relationship between electron and proton scatterings.
Precise determination of not only the values of $r_p$ and $r^2_n$,
but also those of the higher moments$(n>4)$ of the charge
distribution would provide more
information on $\rho_n(r)$ and $\delta R$\cite{ts1}.
For example, Ref.\cite{mark} lists the value of $\avr{r^4}_n$ to be $1317.3$ fm$^4$,
while the LSA in Ref.\cite{ts1} $1282.926$ fm $^4$. 
In comparing $\delta R$s of various experiments with one another,
differences between the estimated values of $\avr{r^2}_p$
should be also examined.

In conclusion, both electron- and proton-scattering experiments are necessary for determination
 of the moments of $\rho_n(r)$.  
The fourth moment, $\avr{r^4}_c$, of the charge
distribution plays a role to verify the consistency 
between the moments of $\rho_\tau(r)$ estimated by those experiments.
Such a role is expected to become more important in future
in studying neutron-rich unstable nuclei\cite{tsu}, where contributions
from the moments of the $\rho_n(r)$
to $\avr{r^n}_c$ would more increase than in stable nuclei\cite{hi}.

\section*{Acknowledgments}
This work was supported by JSPS KAKENHI Grant Numbers JP22K18706.

\theendnotes

\end{document}